

\magnification=\magstep 1
\overfullrule=0pt
\hfuzz=16pt
\voffset=0.0 true in
\vsize=8.8 true in
\baselineskip 20pt
\parskip 6pt
\hoffset=0.0 true in
\hsize=6.5 true in
\nopagenumbers
\pageno=1
\footline={\hfil -- {\folio} -- \hfil}

\ \hfill {\tt Phys. Rev. Lett. 86, 5112-5115 (2001)}

\ 
 
\  

\centerline{\bf Indirect Interaction of Solid-State} 

\centerline{\bf Qubits via Two-Dimensional Electron Gas}
 
\ 

\centerline{\frenchspacing Dima Mozyrsky, Vladimir Privman and M. Lawrence Glasser}

\ 
 
\centerline{\sl Department of Physics, Clarkson University, Potsdam, New York 13699--5820}

\  

\centerline {\bf ABSTRACT}

We propose a mechanism of long-range coherent coupling between nuclear spins to be used as qubits in solid-state semiconductor-heterojunction quantum information processing devices. The coupling is via localized donor electrons which in turn interact with the two-dimensional electron gas. An effective two-spin interaction Hamiltonian is derived and the coupling strength is evaluated. We also discuss mechanisms of qubit decoherence and consider possibilities for gate control of the interaction between neighboring qubits. The resulting quantum computing scheme retains all the gate-control and measurement aspects of earlier approaches, but allows qubit spacing at distances of order 100$\,$nm, attainable with the present-day semiconductor device technologies.    

\vfil

\noindent PACS: 73.20.Dx, 71.70.Ej, 03.67.Lx, 76.60.-k

\vfil\eject

Recent technological advances in electronics related to spin polarization [1,2] have boosted experimental and theoretical interest in quantum information science in condensed matter systems, specifically, in semiconductor heterostructures at low temperatures and in high magnetic fields. The solid state implementations of quantum information devices seem to be among the most promising ones, due to possible scalability of the elementary logic gates into more complicated integrated circuits. Several designs for solid state and related spin-based quantum information processors have been suggested [3-8]. Preliminary experiments, involving several quantum bits (qubits), have been carried out or are being contemplated [9-10]. 

Our work stems from the proposals that utilize nuclear or electronic spins as qubits for information processing [3-7]. These are natural choices for qubits because at low temperatures spin states in semiconductors have relatively long decoherence times, sometimes milliseconds or even longer for electronic spins, and seconds for nuclear spins [11-14]. We propose a new mechanism for coupling between two nuclear-spin qubits, combining aspects of two models of quantum information processors, one based on nuclear spins in quantum Hall effect systems [4], and another utilizing the nuclear spins of phosphorus donors in a silicon heterostructure [5]. 

An appealing aspect of Kane's model [5] is a possibly experimentally feasible scheme for reading out the state of the quantum register, i.e., measurement of a nuclear spin, achieved by transferring the nuclear spin polarization to the electronic state, while the later is measured with the use of a single electron transistor.
The model proposed in [4] has a different advantage: unlike [5], the interaction between the nuclear spins is mediated by the two-dimensional (2D) electron gas, and thus is longer ranged due to the highly correlated state of the 2D electron gas in the quantum Hall regime. This opens up possibilities for experimental realization of such quantum information processors, because large separation between spin qubits means greater lithographic dimensions in manufacturing the device.  
The price paid is that the coupling is weak, and therefore the time scales of the  ``gate function'' can be as large as $1\,$s. 
 
In this work we combine the two proposals, thus retaining the measurement and control scheme proposed in [5,7,9] and at the same time allowing larger separations, of order 100$\,$nm, between interacting qubits. The resulting system is thus realizable with the present-day semiconductor technologies. We propose a model where sparsely positioned phosphorus donors are imbedded in a 2D electron gas in the quantum-Hall regime. The localized donor electrons interact via the delocalized 2D electrons and thus indirectly mediate nuclear-spin interactions. In 3D, spin coupling mechanisms via conduction electrons have been well studied [15]. Here we estimate the range of this induced nuclear-spin interaction for the 2D case and find it to be of order $100\,$nm. This is large compared to atomic dimensions, donor-electron bound state radii, and even the electronic magnetic length which is typically of order $10\,$nm. We find that this interaction is also stronger, thus corresponding to faster ``gate function'' times, than in [4].

We assume that the coupling between the electronic and nuclear donor spins is given by the Fermi contact interaction, $H_{e-n} = A {\bf \sigma}_n \cdot {\bf \sigma}_e$. Here $A=(8\pi/3)\mu_B g_n \mu_n |\Psi_0(0)|^2$, where $\mu_n$ and $g_n$ are the nuclear magneton and nuclear $g$-factor, respectively, $|\Psi_0(0)|^2$ is the donor electron probability density at the nucleus, $\mu_B$ denotes the Bohr magneton, and $\sigma$'s are Pauli matrices. Coupling of the delocalized electrons to the nuclear spin is considerably weaker than that of the localized donor electron. Therefore, we assume that the nuclear spin interacts with conduction electrons indirectly via the donor electron. 

As a prototype system, we consider $^{31}$P donors positioned in Si, so all the spins involved are $1/2$. The donor electronic and nuclear spins form a four-level system. The spectrum of this two-spin system can be obtained to O($A$) with $H_{e-n}$ treated as perturbation. The energy levels are $E_0 = -(\gamma_n + \Delta)/2 + A$, $E_1 = (\gamma_n - \Delta)/2 - A$, $E_2 = (-\gamma_n + \Delta)/2 - A$, and $E_3 = (\gamma_n + \Delta)/2 + A$, where $\gamma_n = g_n \mu_n H$ is the nuclear spin splitting. Here $H$ is the magnetic field, and the expression for $\Delta$, the electronic Zeeman gap, will be given shortly.
The eigenstates associated with these energy levels are $|0\rangle = |\downarrow_e \downarrow_n \rangle $, $|1\rangle = |\downarrow_e \uparrow_n \rangle + (2A/\Delta)|\uparrow_e \downarrow_n \rangle$, $|2\rangle = |\uparrow_e \downarrow_n \rangle - (2A/\Delta)|\downarrow_e \uparrow_n \rangle$, and $|3\rangle = |\uparrow_e \uparrow_n \rangle$, where $|\downarrow_e \downarrow_n \rangle$ represents the electronic and nuclear spin down state, etc. 
Here we propose to consider the states $|0\rangle$ and $|1\rangle$ as qubit states of a quantum computer. By altering the hyperfine coupling constant $A$ by distorting the spatial state $\Psi_0$ of the donor electron with an electrostatic gate [5,7], one can selectively control the state of an individual qubit by means of the NMR technique.

In order to calculate the interaction Hamiltonian between two qubits, we first consider the coupling between the donor electron and conduction electrons.      
The ground state of the donor electron is bound (localized) and will typically lie in the energy gap, several meV below the conduction band edge. For temperatures of order mK, electronic transitions from this localized state to the conduction band are highly improbable. 
The dominant interaction between the localized electron and conduction electrons is their Coulomb interaction. We are interested only in the exchange part of this interaction, i.e., the spin-dependent part. The spin-independent part causes screening, but it is weak in 2D [16] and, especially in the presence of the magnetic field, cannot ionize the donor. 

In large magnetic field, the delocalized 2D electrons occupy highly degenerate Landau energy levels [16]. It is convenient to introduce electron bound state creation and annihilation operators $b^{\dagger}_{n s}$ and $b_{n s}$, where $n$ represents the donor spatial state, and $s$ is the spin $z$-component, $\uparrow$ or $\downarrow$. Let $a^{\dagger}_{m k_x s}$, $a_{m k_x s}$ denote the creation and annihilation operators for the delocalized 2D electrons, where $m$ labels the Landau level, while $\hbar k_x$ is the $x$-momentum (we use the asymmetric gauge). Then the exchange coupling between the bound and delocalized electrons can be written as
{}
$$H_{\rm ex} = {1 \over 2}\sum G^{n,n^{\prime}}_{m,m^{\prime},k_x,k_x^{\prime}} b^{\dagger}_{n s} a^{\dagger}_{m k_x s^{\prime}} b_{n^{\prime} s^{\prime}} a_{m^{\prime} k_x^{\prime} s}\eqno (1)$$ 
{}              
\noindent where the sum is over all the indices. Here we have neglected the spin-orbit interaction. In what follows, we will retain only the lowest donor electron spatial state, i.e., account only for the transitions between the two Zeeman levels of the ground state. 

The 2D electrons are assumed to be in a nondissipative quantum Hall state with filling factor $\nu = 1$, i.e., the lower Zeeman sublevel of the Landau ground state is completely filled [4]. This choice ensures reduced decoherence and relaxation effects [14], owing to the energy gap in the spectrum of the lowest-energy spin-wave excitations which are well studied [17,18]; their spectrum is given by ${\cal E}_{\bf k} = \Delta + E_c \left[ 1 - I_0\left(\ell^2 k^2 / 4\right) \exp\left(-\ell^2 k^2 /4\right)\right] $, where $I_0$ is the modified Bessel function.  
Here $\Delta = g \mu_B H$ is the Zeeman gap, $E_c = \left({\pi / 2}\right)^{1/2} \left({e^2 / \epsilon \ell}\right)$ is the characteristic Coulomb energy, and $g$ is the effective $g$-factor in the potential well that holds the 2D electron gas, while $\epsilon$ is the dielectric constant of the material, and $\ell = (\hbar c /eH)^{1/2}$ is the magnetic length. Extension to larger integer filling factors is possible [14,17,18]. One can also introduce [18] normalized creation and annihilation operators for the spin waves, quadratic in electronic operators, 
{}
$$ S_{\bf k}^{\dag} = \left({2\pi \ell^2 \over L_x L_y}\right)^{1 / 2}\, \sum_p \, e^{i\ell^2 k_y p}  \, a^{\dag}_{p+{k_x \over 2},\downarrow} \, a_{p-{k_x \over 2},\uparrow}\eqno (2)$$
{}
\noindent Here $L_{x,y}$ are the transverse dimensions, taken to infinity in the final calculation. The summation over $p$ is taken in such a way [18] that the wave number subscripts are quantized in multiples of $2\pi /L_x$. The spectrum of these spin-waves has been experimentally verified in GaAs heterostructures [19].      

We will include only these lowest excitations in the sum (1); our goal is to rewrite (1) in terms of the spin-wave operators (2). The exchange coupling is thus truncated to $G^{n,n^{\prime}}_{m,m^{\prime},k_x,k_x^{\prime}} = G_{k_x,k_x^{\prime}} \delta_{n,0}   
\delta_{n^{\prime},0} \delta_{m,0} \delta_{m^{\prime},0}$, where
{}
$$ G_{k_x,k_x^{\prime}} = \int d^3{\bf R}_1 d^3{\bf R}_2 \Psi^{\ast}_0 ({\bf R}_1) \Psi_0({\bf R}_2) U({\bf R}_1 - {\bf R}_2) \Phi^{\ast}_{0,k_x} ({\bf R}_2) \Phi_{0,k_x^{\prime}} ({\bf R}_1), \eqno (3)$$
{}
\noindent $U({\bf R}_1 - {\bf R}_2) = e^2/\epsilon |{\bf R}_1 - {\bf R}_2|$ is the Coulomb interaction and $\Psi_0({\bf R})$ is the donor electron ground state.  The states of the conduction electrons confined in the 2D well are $\Phi_{0,k_x^{\prime}}({\bf R}) = \phi_{0,k_x^{\prime}}({\bf r}) \chi (z)$, where $\phi_{0,k_x^{\prime}}({\bf r})$ are the standard 2D Landau states [16]; $\chi(z)$ describes the confinement of the conduction electron wavefunction in the $z$ direction and depends on the nature of the confinement potential. Here and in the following ${\bf R} = ({\bf r},z)$, with ${\bf R}$ and ${\bf r}=(x,y)$ being 3D and 2D coordinates, respectively, while $z$ is the direction perpendicular to the heterostructure, in which the applied magnetic field is pointing.

With the use of the expressions for Landau ground state wavefunctions, $\phi_{0,k_x^{\prime}}({\bf r})=\ell^{-1}(\pi L_x)^{-1 / 2} e^{i k_x x} \exp[-(y-\ell^2 k_x)^2/2\ell^2]$, and (2), after a lengthy calculation, we get
{}
$$H_{ex}= {1 \over 2} \sum_{\bf k} \left[ W_{\bf k} |\uparrow_e\rangle\langle\downarrow_e| 
S_{\bf k} + W^{\ast}_{\bf k} |\downarrow_e\rangle\langle\uparrow_e|
S^{\dagger}_{\bf k}\right] \eqno (4)$$
{}
\noindent where $|\uparrow_e\rangle\langle\downarrow_e| =  b^{\dagger}_{\uparrow} 
b_{\downarrow}$ in the appropriate subspace, and
{}
$$W_{\bf k} = {1 \over \ell (2\pi L_x L_y)^{1 / 2}} \int d^3{\bf R}_1 d^3{\bf R}_2 \Psi^{\ast}_0 ({\bf R}_1) \Psi_0({\bf R}_2) U({\bf R}_1 - {\bf R}_2) \chi^{\ast} (z_2) \chi (z_1)C_{\bf k} ({\bf r}_1,{\bf r}_2) \eqno (5)$$
{}
$$C_{\bf k} ({\bf r}_1,{\bf r}_2) \equiv \exp{\left\{-{1 \over 4\ell^2}\left[(x_1-x_2)^2 + (y_1-y_2)^2
-2i(x_1-x_2)(y_1+y_2)\right]\right\} } \times$$
$$\exp{\left[ - {\ell^2 \over 4}(k_x^2 + k_y^2) - {k_y \over 2}(iy_1 + iy_2 -x_1+x_2) - {k_x \over 2}(ix_1 + ix_2 + y_1 - y_2)\right]} \eqno (6)$$
{}
\noindent Note that since all the position vectors $\bf R$, $\bf r$ are measured from the origin at the donor atom, the quantity $W_{\bf k}$ depends also on the donor coordinates. To the leading order, (4) gives the interaction of the donor electron spin with excitations of the 2D electron gas in the $\nu = 1$ integer quantum-Hall state.

One can rewrite the interaction (4)-(6), with (4) multiplied by the unit operator in the nuclear-spin Hilbert space, in terms of the eigenstates of the electron-nucleus system. With the use of the expressions derived earlier for these eigenstates in terms of direct products of electronic and nuclear spin states, we obtain 
{} 
$$H_{ex} = {1 \over 2} \sum_{\bf k} W_{\bf k}\left({2A \over \Delta} |1 \rangle\langle 0| + |2 \rangle\langle 0|  + |3 \rangle\langle 1|- {2A \over \Delta}|3 \rangle\langle 2|\right) S_{\bf k} + h.c. \eqno (7)$$
{}
\noindent Now one can calculate an effective Hamiltonian for the interaction of two qubits. Since the electronic
Zeeman gap is much larger than the nuclear one, we can truncate the Hilbert space of the combined electron-nucleus spins to the two lowest lying states. Thus we retain only the $|0 \rangle\langle 1|$ and conjugate transitions in the exchange interaction (7). 

An effective interaction between two qubits can be obtained within the standard framework of second order perturbation theory by tracing out the states of the spin waves; see [15,20,21] for similar calculations. 
The result can be written as 
{}
$$H_{1,2} = J |0_1 1_2 \rangle\langle 1_1 0_2| + J^{\ast} |1_1 0_2 \rangle\langle 0_1 1_2| \eqno (8)$$
{}
\noindent Here the coupling constant between the two qubits is
{}
$$ J = \left({A \over \Delta}\right)^2 \sum_{{\bf k}\neq 0} \, {W_{{\bf k},1} W^{\ast}_{{\bf k},2} \over {\cal E}_{\bf k}+E_1-E_0} \eqno (9)$$ 
{}
\noindent The subscripts 1 and 2 in (8)-(9) label the two donor qubits, while $W_{{\bf k},1}$ and $W_{{\bf k},2}$ are the coupling constants of each donor electron spin to spin waves, given by (5), and ${\cal E}_{\bf k}$ is the spin-wave energy. 

The nuclear-spin energy gap is much smaller
than the electronic spin-wave excitation energies. Therefore, 
we can ignore $E_1 - E_0$ in the denominator in (9).
Furthermore, due to the large value of the spin-wave
spectral gap at ${\bf k} = 0$, ${\cal E}_0 = \Delta$,
we do not have the ``small denominator'' problem encountered
in other calculations of this sort, e.g., [20].
Physically this means that the spin excitations in the 2D electron gas mediating the effective qubit-qubit interaction are virtual and so this interaction does not cause appreciable relaxation or decoherence on the ``gate function'' timescale $\hbar/J$. 

It is important to note that one can construct a universal CNOT logic gate from the controlled dynamics governed by Hamiltonians of the form of $H_{1,2}$ and single qubit rotations [6]. The coupling strength $J$ between the qubits can be externally controlled by the electrostatic gates built above the 2D inversion layer. By applying gate voltages, one can locally vary the density of the 2D electrons thus changing coupling between the delocalized and donor electrons. This results in control over the effective coupling constant $J$ in (9). The precise effect of gates on interactions between the qubits as well as on decoherence of their states, should be further studied in order to establish the feasibility of the quantum-computing approach proposed here. Most other semiconductor solid-state quantum-computing approaches [3-7] utilize gates.

Let us explicitly calculate the coupling constant $J$ in (8)-(9). Because the spatial ground state of the donor is localized on a scale smaller than the magnetic length $\ell$, the overlap integrand in (5) is vanishingly small for $|{\bf r}_1 - {\bf r}_2| > \ell$. At the same time, for $|{\bf k}| > 1/\ell$, the value of $C_{\bf k}$ decreases exponentially. Thus $C_{\bf k}$ can be simplified by neglecting $x_1 - x_2$ and $y_1 - y_2$ terms in (6). Moreover, for two donors at separation larger than $\ell$, we can put $({\bf r}_1 + {\bf r}_2)/2 \simeq {\bf r}_j$, with ${\bf r}_j$ being the location of either one of them. Then (5) can be approximated by $W_{{\bf k},j} = Z(L_x L_y)^{-1 / 2} \exp\left(-{\ell^2 k^2 \over 4} - i{\bf k} \cdot{\bf r}_{j}\right)$, 
with $Z = \left(1 / 2\pi\ell^2\right)^{1 / 2} \int d^3{\bf R}_1 d^3{\bf R}_2 \Psi^{\ast}_0 ({\bf R}_1) \Psi_0({\bf R}_2) U({\bf R}_1 - {\bf R}_2) \chi^{\ast} (z_2) \chi (z_1)$. 

Finally, the coupling constant $J$ of the effective interaction (8) can be obtained by transforming the summation in (9) to integration in the limit $L_{x,y} \to \infty$,
{}
$$J = \left({A \over \Delta}\right)^2{|Z|^2 \over (2\pi)^{1 / 2} E_c \ell^2} \left({d \over r}\right)^{1 / 2} \exp\left(-{r \over d} \right) \; , \; \qquad (r > \ell) \eqno (10)$$ 
{}
\noindent where $d=\left(E_c/2\Delta\right)^{1 / 2}\ell$. A similar dependence of the coupling on the donor separation $r$ was obtained in a study of nuclear polarization diffusion in the quantum Hall regime [21]. Interaction (7)-(8) between the spins has finite range $d$, which, however, is very large compared to the effective Bohr radius of the donor ground state. Thus the indirect exchange at large distances dominates the direct exchange interaction resulting from the overlap of the two atomic wavefunctions. For magnetic field $H=6\,$T and $\epsilon=12$, we get $d\simeq 65\,$nm, which is indeed much greater than the characteristic Bohr radius for a donor electron in silicon. 

In order to estimate $J$, we have to evaluate the overlap integral $Z$. For an order-of-magnitude estimate we will assume that $\chi(z)$ is constant inside the well and zero outside.
Then $Z \simeq (2\pi)^{-1 / 2} (\delta\ell)^{-1} \int d^3{\bf R}_1 d^3{\bf R}_2 \Psi^{\ast}_0 ({\bf R}_1) \Psi_0({\bf R}_2) U({\bf R}_1 - {\bf R}_2)$, where $\delta$ is the width of the well. We put $\delta \simeq 4\,$nm. For $\Psi_0({\bf R})$, the donor ground state, we choose a spherically symmetric Hydrogen-like ground state with the effective Bohr radius $a_B \simeq 2\,$nm. This is, of course, not the case in a realistic situation [22]. The ground state of the donor will be influenced by the band structure, by the magnetic field and by the confining 2D well potential, while the states of the conducting electrons will be distorted by the impurity potentials. We are not aware of a thorough study of these effects for our system. For the purposes of an order of magnitude estimate, however, a spherical state should be sufficient. 

Evaluating the integral for the Coulomb potential $U$, we obtain $Z \simeq (5a^2_B/16\delta) E_c$. Assuming that the two donors are separated by the distance $r=100\,$nm and using the value $2A/h = 58 \,$MHz from [4], we obtain the estimate $J/\hbar \sim 10^2 \,$s$^{-1}$. 

The clock speed of the information processor just described appears to be a fraction of kHz and should be compared with the time scales for relaxation and decoherence.
The leading mechanism for these at low temperatures is through interaction with impurities. It has been found theoretically [12,23] and confirmed experimentally [2] that nuclear spin relaxation in the quantum Hall regime is slow and strongly dependent on the impurity potentials; typically, the relaxation time $T_1$ is of order $10^3 \,$s. In our case the interaction of a qubit with the 2D gas is stronger, and as a result, the relaxation is expected to be faster. An estimate from formulas in [12,23] gives $T_1 \simeq 1 \,$s. There is, however, another 
important issue---decoherence, on time scales $T_2$. Recently, this quantity has been calculated in the same framework, that is, when the interaction of the conduction electrons with impurities is taken into account [14]. The results of [14] can be adjusted for the present case and yield the estimate $T_2 \simeq 10^{-1} \,$s.

The existing quantum error correction protocols require the quality factor, equal the ratio of the gate-function clock time to decoherence time, not to exceed $10^{-5}$ [24]. Our estimates indicate that this is not the case for the present system. Actually, no quantum computing proposal to date, scalable by other criteria, satisfies this $10^{-5}$ quality-factor criterion. The values range from $10^{-1}$ to $10^{-3}$. The resolution could come from development of better error-correction algorithms or from improving the physical system to obtain a better quality factor. In our estimate of the decoherence time scale, we used parameters typical of a standard, ``dirty'' heterostructure with large spatial fluctuations of the impurity potential. These heterostructures have been suitable for standard experiments because they provide wider quantum-Hall plateaus. Much cleaner, ultra-high mobility structures can be obtained by placing the ionized impurity layer at a larger distance from the 2D gas or by injecting conduction electrons into the heterostructure by other means. 

Thus, our present quantum-computing proposal offers a clear direction for exploring a physical, rather than algorithmic, resolution to the quality factor problem. This possibility should be further examined both experimentally and theoretically. 
Our new quantum computing paradigm suggests several interesting avenues for research. The effect of gates on the switching of qubit interactions and on decoherence requires further investigation. The first experimental realizations will probably involve only few qubits. The interactions of these may be significantly affected by the geometry, specifically, the edges, of the heterostructure.
 
The authors acknowledge useful discussions with I.\ D.\ Vagner. This research was supported by the National Security Agency (NSA) and Advanced Research and Development Activity (ARDA) under Army Research Office (ARO) contract number DAAD$\,$19-99-1-0342.

\vfill\eject

\centerline {\bf References}

{\frenchspacing

\vskip 0.12 in
 
\item{1.} R. Tycko, S.E. Barrett, G. Dabbagh, L.N. Pfeiffer and K.W. West, Science {\bf 268}, 1460 (1995); S.E. Barrett, G. Dabbagh, L.N. Pfeiffer, K.W. West and R. Tycko, Phys. Rev. Lett. {\bf 74}, 5112 (1995).

\item{2.} A. Berg, M. Dobers, R.R. Gerhards and K.v. Klitzing, Phys. Rev. Lett. {\bf 64}, 2563 (1990).

\item{3.} D. Loss and D.P. DiVincenzo, Phys. Rev. A {\bf 57}, 120 (1998).

\item{4.} V. Privman, I.D. Vagner and G. Kventsel, Phys. Lett. A {\bf 239}, 141 (1998).

\item{5.} B.E. Kane, Nature {\bf 393}, 133 (1998).  

\item{6.} A. Imamoglu, D.D. Awschalom, G. Burkard, D.P. DiVincenzo, D. Loss, M. Sherwin and A. Small, Phys. Rev. Lett. {\bf 83}, 4204 (1999).

\item{7.} R. Vrijen, E. Yablonovitch, K. Wang, H.W. Jiang, A. Balandin, V. Roychowdhury, T. Mor and D.P. DiVincenzo, Phys. Rev. A {\bf 62}, 012306 (2000).

\item{8.} Yu. Makhlin, G. Schoen and A. Shnirman, Nature {\bf 398}, 305 (1999); P.M. Platzman and M.I. Dykman, Science {\bf 284}, 1967 (1999); N.A. Gershenfeld and I.L. Chuang, Science {\bf 275}, 350 (1997).

\item{9.} B.E. Kane, N.S. McAlpine, A.S. Dzurak, R.G. Clark, G.J. Milburn, H.B. Sun and H. Wiseman, Phys. Rev. B {\bf 61}, 2961 (2000).

\item{10.} D.G. Cory, A.F. Fahmy and T.F. Havel, Proc. Natl. Acad. Sci. USA {\bf 94}, 1634 (1997); J.A. Jones and M. Mosca, J. Chem. Phys. {\bf 109}, 1648 (1998); J.R. Tucker and T.-C. Shen, Solid State Electronics {\bf 42}, 1061 (1998); R.G. Clark, unpublished results presented at a quantum-computing conference held in Baltimore, MD, August 2000.

\item{11.} D.P. DiVincenzo, Science {\bf 270}, 255 (1995).

\item{12.} I.D. Vagner and T. Maniv, Phys. Rev. Lett. {\bf 61}, 1400 (1988).

\item{13.} Y. Yaffet, in {\it Solid State Physics}, edited by F. Seitz and D. Turnbull, Vol. {\bf 14\/} (Academic Press, New York, 1963); H. Hasegawa, Phys. Rev. {\bf 118}, 1523 (1960); L.M. Roth, Phys. Rev. {\bf 118}, 1534 (1960).

\item{14.} D. Mozyrsky, V. Privman and I.D. Vagner, Phys. Rev. B 63, 085313 (2001).

\item{15.} T. Kasuya, in {\it Magnetism}, edited by G.T. Rado and H. Suhl, Vol. {\bf 2B} (Academic Press, New York, 1966).

\item{16.} {\it The Quantum Hall Effect}, edited by R.E. Prange and S.M. Girvin (Springer-Verlag, New York, 1987).

\item{17.} Yu.A. Bychkov, S.V. Iordanskii and G.M. Eliashberg, JETP Lett. {\bf 33}, 143 (1981).

\item{18.} C. Kallin and B.I. Halperin, Phys. Rev. B {\bf 30}, 5655 (1984).

\item{19.} A. Pinczuk, B.S. Dennis, D. Heiman, C. Kallin, L. Brey, C. Tejedor, S. Schmitt-Rink, L.N. Pfeiffer and K.W. West,  Phys. Rev. Lett. {\bf 68}, 3623 (1992).

\item{20.} M.A. Ruderman and C. Kittel, Phys. Rev. {\bf 96}, 99 (1954).

\item{21.} Yu.A. Bychkov, T. Maniv and I.D. Vagner, Solid State Commun. {\bf 94}, 61 (1995).

\item{22.} W. Kohn, in {\it Solid State Physics}, edited by F. Seitz and D. Turnbull, Vol. {\bf 5} (Academic Press, New York, 1957).

\item{23.} S.V. Iordanskii, S.V. Meshkov and I.D. Vagner, Phys. Rev. B {\bf 44}, 6554 (1991).

\item{24.} J. Preskill, Proc. R. Soc. Lond. A {\bf 454}, 385 (1998).
 
}
 
\bye